# An Approximate Solution Method for Large Risk-Averse Markov Decision Processes

**Marek Petrik**
IBM Research
Yorktown Heights, NY 10598
mpetrik@us.ibm.com

**Dharmashankar Subramanian**
IBM Research
Yorktown Heights, NY 10598
dharmash@us.ibm.com

## Abstract

Stochastic domains often involve risk-averse decision makers. While recent work has focused on how to model risk in Markov decision processes using risk measures, it has not addressed the problem of solving large risk-averse formulations. In this paper, we propose and analyze a new method for solving large risk-averse MDPs with hybrid continuous-discrete state spaces and continuous action spaces. The proposed method iteratively improves a bound on the value function using a linearity structure of the MDP. We demonstrate the utility and properties of the method on a portfolio optimization problem.

## 1 Introduction

It is common for decision makers to be risk-averse when uncertainty is involved. For example, risk-averse financial portfolio managers prefer an investment portfolio that mitigates the worst-case plausible losses even if the corresponding expected return is suboptimal. Since the worst-case losses corresponding to solutions that optimize the expected returns can be unacceptable to such decision-makers, there is a need to explicitly model risk-averse objectives. While it is common to use utility functions to capture the decision maker's tolerance of risk, this approach cannot capture many empirically observed risk-averse preferences (e.g. [13]). In addition, decision makers are often unable to provide appropriate utility functions, and non-linear utility functions can complicate the application of dynamic programming [12]. In this paper, we focus on *coherent risk measures*—an alternative model of risk-aversion.

Coherent measures of risk were developed in the field of finance and represent an alternative model of risk-avoidance to utility functions [2, 5]. A coherent risk measure $\varrho$ is a generalization of the risk-neutral expectation operator that maps a random variable $X$ to a real number. Informally, while the risk-neutral expectation computes the average corresponding to a given reference probability distribution, a coherent risk measure takes the worst average assessed over a suitably defined neighborhood around the reference distribution. As such, a coherent risk measure is also interpretable as a conservative way to account for ambiguity in the precise knowledge of the reference distribution. To be a coherent risk measure, the function $\varrho$ must satisfy properties that include convexity and monotonicity, which we describe in detail later.

The properties of coherent risk measures in static settings have been studied extensively and are very well understood [5]. Their application to dynamic settings, in which decisions are taken sequentially as additional stochastic information arrives, is more complicated. Dynamic risk measures have been usually studied in stochastic programming settings which we discuss in Section 6. Recently, dynamic risk measures have been used to formulate risk-averse Markov decision processes [21, 13]. Unfortunately, the methods proposed in these papers only solve small MDPs in which both the states and actions can enumerated. In this paper, we extend the modeling methodology of [21] and [13] to solve structured MDPs with continuous state and action spaces.

To illustrate an MDP with continuous state and action spaces, consider the problem of managing a stock portfolio [11]. In this problem, the decision maker must allocate the investment among a portfolio of stocks in order to maximize the financial gain. The one-step gain of each stock is stochastic, but depends on the current market state; i.e. the distribution of returns is different for volatile and calm markets. The objective is to maximize the expected returns while minimizing the risk of significant losses.

The main components of a Markov decision process are the state and action spaces. For example, the state in the portfolio management problem is represented by the current position in the stocks and cash, and the state of the market. While the current position may be modeled as a continuous vector, it is common to model the market in terms of a finite set of possible market states [11]. The resulting state space is naturally a hybrid discrete-continuous state space. The actions represent the change in the investments in every period and are also continuous. Because of the continuous aspects of the state and action sets, this problem cannot be solved using the methods proposed in [13]. Large MDPs such as this one are typically solved using approximate dynamic programming, a reinforcement learning method [18]. The method that we propose is related to approximate dynamic programming, but it also exploits the linearity of the domain and the coherent risk measure.

The remainder of the paper is organized as follows. First, Section 2 introduces coherent risk measures and describes their properties in multi-stage optimization. Section 3 then defines risk-averse hybrid linearly controlled MDPs. These problems have both continuous and discrete states, continuous actions, linear transitions, and their objective is formulated in terms of a coherent risk measure. In Section 4, we propose Risk-averse Dual Dynamic Programming (RDDP)—a method to solve hybrid risk-averse MDPs. Our method is related to stochastic dual dynamic programming and point-based methods for solving POMDPs [16]. In Section 5, we evaluate the algorithm on the portfolio management problem and analyze the solution properties. Finally, Section 6 discusses the connections to relevant work on risk-averse objectives in multi-stage stochastic programs.

## 2 Risk Measures in Markov Decision Processes

In this section, we briefly overview the state of the art in applying risk measures in dynamic settings. We formally define risk-averse Markov decision processes and other necessary notation. We omit some minor technical assumptions needed for continuous state and action sets; refer for example to Sections 4.3 and 4.4 of [19] for the technical details.

A *Markov decision process* is a tuple $(\mathcal{S}, \mathcal{A}, W, P, c)$ defined for time steps $\mathcal{T} = \{0 \ldots T\}$ as follows. Let $\mathcal{S}$ be a *compact* state space and $\mathcal{A} \subseteq \mathbb{R}^m$ be a *compact* continuous action space. Let $W : \mathcal{S} \rightrightarrows \mathcal{A}$ denote a measurable set-valued function (or a multimap) that maps each state to the *compact* set of permissible actions and let $P : \mathcal{S} \times \mathcal{A} \to \mathcal{P}(\mathcal{S})$ denote the transition probabilities where $\mathcal{P}(\mathcal{S})$ is the set of all probability measures on $\mathcal{S}$; $P(s, a)$ represents the next-state probability measure for any $s \in \mathcal{S}$ and $a \in \mathcal{A}$. Let $c : \mathcal{S} \times \mathcal{A} \times \mathcal{S} \mapsto \mathbb{R}$ represent the transition cost; $c(s_t, a_t, s_{t+1})$ represents the cost incurred when transiting from $s_t$ to $s_{t+1}$ upon taking action $a_t$. Finally, let $s_0$ denote a deterministic initial state. In the final time-step $T + 1$, we assume that there is no action taken and the problem terminates.

The focus of this paper is on *finite*-horizon total return models. The solution of a finite-horizon MDP is a policy which is composed of decision rules. A *decision rule* $d_t : \mathcal{S} \to \mathcal{A}$ at time $t$ determines the action to be taken in each state. A collection of decision rules, one per each time step, is a deterministic Markov *policy* $\pi : \mathcal{T} \times \mathcal{S} \mapsto \mathcal{A}$. The set of all policies is denoted as $\Pi$.

The objective in solving risk-neutral MDPs is to minimize the total cost, or *return*, which is defined for $\pi = (d_1, \ldots, d_t)$ as:

$$\varsigma(\pi) = \mathbf{E}_{P_\pi} \left[ \sum_{t \in \mathcal{T}} c(S_t, d_t(S_t), S_{t+1}) \right], \qquad (1)$$

where $S_t$ is a random variable that represents the state at time $t$, distributed according to the probability distribution $P_\pi$ induced by the policy $\pi$. The optimal return is $\varsigma^\star = \min_{\pi \in \Pi} \varsigma(\pi)$. It is well-known that there exists an optimal Markov deterministic policy, so it suffices to restrict our attention to this class of policies.

Risk averse MDPs have been formalized recently in [21] and [13]. Informally, given a reference probability distribution for each transition, these formulations modify the objective Eq. (1) by specifically penalizing a subset of the most adverse realizations of the induced cost distribution. In particular, the expectation is replaced by a *dynamically consistent* risk measure $\varrho$. Risk measures that are dynamically consistent can be applied to multistage optimization problems without violating the dynamic programming principle [22].

Assume that the one-step random MDP costs corresponding to action $a_t$ taken at time step $t$ are represented by $\{Z_{t+1}\}_{t \in \mathcal{T}}$ where $Z_{t+1} \in \mathcal{Z}_{t+1}$, and $\mathcal{Z}_{t+1}$ is the space of random variables adapted to the information filtration at time $t+1$, i.e. the history of the underlying stochastic process at time $t+1$. An ordinary risk measure would assign the objective based on the total sum of the costs $\varrho(\{Z_{t+1}\}_{t \in \mathcal{T}}) = \rho(\sum_{t \in \mathcal{T}} Z_{t+1})$. A dynamically consistent risk measure, on the other hand, is defined as a composition of one-step risk mappings:

$$\varrho(Z_1, \ldots, Z_{T+1}) = \rho_0 \left( Z_1 + \rho_1 \left( Z_2 + \ldots + \rho_T \left( Z_{T+1} \right) \right) \right).$$

The one-step conditional risk mappings $\rho_t : \mathcal{Z}_{t+1} \to \mathcal{Z}_t$ must satisfy the following properties:

(A1) *Convexity:* $\rho_{t+1} (\alpha \cdot Z + (1-\alpha) \cdot Z') \leq \alpha \cdot \rho_{t+1}(Z) + (1-\alpha) \cdot \rho_{t+1}(Z')$ for all $Z, Z' \in \mathcal{Z}_{t+1}$ and $\alpha \in [0, 1]$.

(A2) *Monotonicity:* If $Z \preceq Z'$ then $\rho_{t+1}(Z) \preceq \rho_{t+1}(Z')$.
(A3) *Translation equivariance:* If $a \in \mathcal{Z}_t$, then $\rho_{t+1}(Z + a) = \rho_{t+1}(Z) + a$.
(A4) *Positive homogeneity:* If $t > 0$, then $\rho_{t+1}(t \cdot Z) = t \cdot \rho_{t+1}(Z)$.

Note that the value $\rho_t(Z)$ is a random variable in $\mathcal{Z}_t$ and the inequalities between random variables are assumed to hold almost surely.

A general dynamically consistent risk measure may be cumbersome to use due to two reasons. The parameters that specify the particular choice of the mapping $\rho_t$ could in general depend on the entire history until time $t$, and further, the argument on which the mapping applies, namely, $Z_{t+1}$ could in general depend on the entire history until time $t+1$, therefore requiring non-Markov optimal policies. We, therefore, further restrict our treatment to *Markov risk measures* [21].

Markov risk measures require that the parameters that specify the particular choice of the conditional risk mappings $\rho_t$ are independent of the history, given the current state, and further that the mapping operation is also independent of the history, given the current state. This is a natural choice for MDPs because the source of uncertainty is the stochastic transition corresponding to any given state-action pair $(s_t, a_t)$, and the reference probability distribution for this stochastic transition is independent of the history, given the current state. Intuitively, an input to a conditional risk mapping of a Markov risk measure $\rho_t$ is the random variable of the costs incurred during a transition from a state $s_t \in \mathcal{S}$ and the output of the risk mapping is a single number for each $s_t$.

The simplest example of a conditional risk mapping is the expectation in which case the objective becomes risk-neutral. The risk neutral objective would be:

$$\varrho(Z_1, \ldots, Z_{T+1}) = \mathbf{E}[Z_1 + \mathbf{E}[\ldots + \mathbf{E}[Z_{T+1}]]].$$

A more general risk measure that allows a convenient specification of the appropriate level of risk tolerance is the Average Value at Risk (AV@R$_\alpha$) parameterized at a chosen level $\alpha > 0$ which is defined as [5]:

$$\text{AV@R}_\alpha(Z) = \max_{q \in \mathcal{Q}_\alpha} \mathbf{E}_q[Z], \quad (2)$$

$$\mathcal{Q}_\alpha = \left\{ q : q(\omega) \leq \frac{p(\omega)}{\alpha}, 0 \leq q(\omega) \leq 1, \omega \in \Omega \right\}.$$

Here, $\Omega$ is the *finite* sample space and $p$ is a reference distribution which, in our case, is the distribution induced by the transition probabilities. It is easy to see that AV@R$_1(Z) = \mathbf{E}[Z]$ and AV@R$_0(Z)$ is the worst-case (or robust) realization. In plain words, average value at risk is the conditional expectation beyond the $\alpha$ quantile. Average value at risk, therefore, will assign a higher overall cost to a scenario with heavier tails even if the expected value stays the same. Fig. 1 illustrates how a AV@R is computed in comparison with a plain expectation.

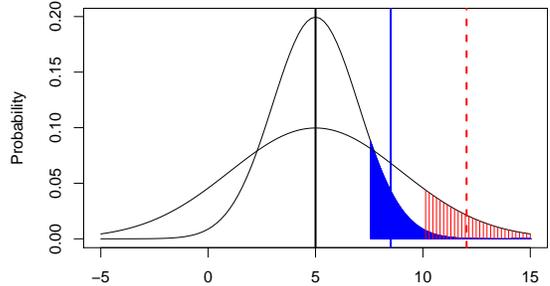

Figure 1: Comparison of two distributions with identical expectations and different AV@R$_{0.1}$ values. The filled-in regions illustrate the quantiles, while the vertical lines indicate the expectations and values at risk.

In the interest of clarity, we concentrate in the remainder of the paper on conditional risk mappings that are a convex combination of average value at risk and expectation and defined as follows:

$$\rho_t(X) = (1 - \lambda)\mathbf{E}_p[X] + \lambda \text{AV@R}_\alpha(X) \quad (3)$$

We will use the fact that the set $\mathcal{Q}_\alpha$ is polyhedral to derive the algorithms. However, it is well known that any conditional risk mapping that satisfies the assumptions (A1)-(A4) can be represented as:

$$\rho_t(Z) = \sup_{Q \in \mathcal{Q}(s_t, a_t)} \mathbf{E}_Q[Z]$$

where $\mathcal{Q}(s_t, a_t) \subseteq \mathcal{P}(\mathcal{S})$ is a closed bounded convex set of probability measures on $\mathcal{S}$. In addition, when the conditional risk mapping satisfies the *comonotonicity* assumption [5], the set $\mathcal{Q}$ is polyhedral. The proposed algorithm can be, therefore, applied to any dynamically consistent risk measure in which the conditional risk mappings satisfy the comonotonicity assumption.

To derive RDDP, we need yet another representation of AV@R as a minimization linear program. Because the set $\mathcal{Q}_\alpha$ is a polytope, the optimization in Eq. (3) can be written as a linear program. The dual of this linear program yields the following representation:

$$\rho_t(Z) = (1-\lambda)\mathbf{E}_p[Z] + \lambda \begin{bmatrix} \min_{\mu, \xi} & \mu + \frac{1}{\alpha} p^\mathsf{T} \xi \\ \text{s.t.} & \xi + \mathbf{1}\mu \geq Z \\ & \xi \geq \mathbf{0} \end{bmatrix} \quad (4)$$

The duality we used above for finite spaces can be generalized to more general sample spaces under reasonable technical assumptions [20]. The optimal value of $\mu$ corresponds to the quantile of $Z$ at level $\alpha$.

One attractive property of dynamically consistent risk measures is that they not only model risk aversion, but also preserve most of the structure of MDPs that makes them easy to solve. In particular, there exists an optimal deterministic Markov policy in risk-averse MDPs with Markov risk measures [21], under some mild additional technical assumptions. The additional technical conditions that are needed are the continuity of $P(s_t, \cdot)$ and lower semicontinuity of $W$ and $c(s_t, a_t, s_{t+1})$ with respect to $a_t$.

It is also possible to define the *value function* in risk-averse MDPs with Markov risk measures. The finite-horizon value function $v_t : \mathcal{S} \to \mathbb{R}$ is defined identically as for the risk neutral case, except the expectations are replaced by a risk measure; that is for some policy $\pi$ the value function represents the return obtained when starting in a given state. The action-value function $q_\pi : \mathcal{T} \times \mathcal{S} \times \mathcal{A} \mapsto \mathbb{R}$ for a policy $\pi$, also known as $Q$-function [1, 18], represents the value function in a state $s$ *after* taking an action $a$. The optimal value function is a value function that corresponds to an optimal policy.

The optimal value function in risk-averse MDPs must satisfy the Bellman optimality conditions, which are stated in the following theorem.

**Theorem 1** ([21]). *A finite-horizon value function $v^\star(t)$ is optimal if and only if it satisfies for all $s \in \mathcal{S}$ and $t \in \mathcal{T}$ the following equality:*

$$v_t^\star(s_t) = \min_{a \in \mathcal{A}} q_t^\star(s_t, a)$$
$$q_t^\star(s_t, a) = \rho_t \left( c_t(s_t, a, S_{t+1}) + v_{t+1}^\star(S_{t+1}) \right).$$

*Here, $S_{t+1}$ is a random variable representing the state at $t+1$ distributed according to the transition probability $P(s_t, a)$ and $v_{T+1}(s) = 0$.*

We will use RDDP to compute an approximately optimal value function for a risk averse MDP. The actual solution is the greedy policy with respect to this value function. A policy $\pi$ is *greedy* with respect to a value function $q$ when it is defined for any $s_t \in \mathcal{S}$ at time $t$ as:

$$\pi(s_t) \in \arg\min_{a \in \mathcal{A}} q_t(s_t, a) \ .$$

Using the greedy policy is the most common method of computing a policy from a value function.

This section introduced a very general model for risk-averse Markov decision processes. In the remainder of the paper, Section 3 defines a linear structure of the state and action spaces that is common in some domains, and Section 4 derives RDDP that exploits this structure to efficiently approximate the optimal value function.

## 3 Hybrid Linearly Controlled Problems

Standard MDP solution methods, such as value or policy iteration, do not scale easily to large or continuous Markov decision processes. Unfortunately, many practical applications involve MDPs with continuously many states and actions. In this section, we describe a specific class of risk averse MDPs with linear transitions and costs that can be solved more efficiently.

Informally, a hybrid linearly controlled problem is an MDP with a state set that has both a discrete and a continuous component. The actions are continuous. The transition probabilities are independent of continuous states and the transitions for continuous states can be described linearly in terms of the continuous states and actions. The following definition describes the problem formally.

**Definition 1.** *A hybrid linearly controlled problem is an MDP $(\mathcal{S}, \mathcal{A}, W, P, c)$ such that:*

- *States $\mathcal{S} = \mathcal{D} \times \mathcal{X}$ where $\mathcal{X} \subseteq \mathbb{R}^n$ is a closed polyhedral set of continuous states and $\mathcal{D}$ is a (small) finite set of discrete states.*
- *Actions $\mathcal{A} \subseteq \mathbb{R}^m$ is a closed convex polyhedral set for some $m$.*
- *Admissible actions for $(d, x) \in \mathcal{S}$ are restricted by the set-valued map $W : \mathcal{S} \rightrightarrows \mathcal{A}$ defined as:*

  $$W(d, x) = \{a \in \mathcal{A} \ : \ A_d a = b_d - X_d x, l_d \le a \le u_d\}.$$

- *Transitions $P : \mathcal{S} \times \mathcal{A} \to \mathcal{P}(\mathcal{S})$ are defined as follows. Assume a finite sample space $\Omega_d$ for each $d \in \mathcal{D}$ with a given reference probability distribution $P_D$. $\Omega_d$ can be seen as a $d$-specific, one-step finite sample space corresponding to the evolution of an underlying exogenous, stationary, stochastic process. Further, assume the following random variables are specified: $D : \Omega_d \to \mathcal{D}, T_x : \Omega_d \to \mathbb{R}^{n \times n}, T_a : \Omega_d \to \mathbb{R}^{n \times m}, U : \Omega_d \to \mathbb{R}^n$. The mapping $D$ captures the action-independent transition from one discrete state to another in the set $\mathcal{D}$. Now define a random variable:*

  $$X = T_x x + T_a a + U \ .$$

  *The mapping $X$ captures the transition from one continuous state to another in the set $\mathcal{X}$ upon taking action $a$. Both the discrete and continuous state transitions are induced by $P_D$. In other words, $P((d, x), a)$ is the probability distribution induced by $P_D$ over all (random) pairs $(D, X)$.*

- *Cost function $c : \mathcal{S} \times \mathcal{A} \times \mathcal{S} \to \mathbb{R}$ is a linear function in terms of states and actions:*

  $$c(s, a, s') = c((d, x), a, (d', x'))$$
  $$= c_a^\mathsf{T} a + c_x^\mathsf{T} x + c_n^\mathsf{T} x'.$$

For sake of simplicity, we assume that there exists an admissible action for every state ($W$ is non-empty). This is equivalent to the complete recourse assumption, which is common in stochastic programming.

**Dynamic Portfolio Management**

We now use a portfolio optimization problem—described in [11]—to illustrate the above model of a hybrid linearly controlled system; later we use this model to experimentally evaluate RDDP. The model assumes three risky assets, cash, and a market state. The one-step returns of the risky assets are denoted by a vector $r_t = (r_t^1, \ldots, r_t^3)$ for any time $t$. In addition, there is assumed to be a risk-free asset—cash—with a constant and known return $r^f$. All returns are inflation adjusted. The monetary holdings of the assets at time $t$ are denoted by a vector $x_t = (x_t^1, \ldots, x_t^3)$ and the risk-free cash position is $c_t$. The decision maker decides on trades that are bought (positive) and sold (negative) on a daily basis, which are denoted by $a_t = (a_t^1, \ldots, a_t^3)$. The transaction costs $\kappa(a_t)$ for a trade are proportional and defined as:

$$\kappa(a_t) = \sum_{i=1}^{3} \delta_i^+ \left[a_t^i\right]_+ - \delta_i^- \left[a_t^i\right]_-$$

for some non-negative $\delta_i^+$ and $\delta_i^-$. The asset holdings $x_t$ and cash position $c_t$ evolve according to:

$$x_{t+1}^i = r_{t+1}^i \cdot (x_t^i + a_t^i) \quad \forall i$$
$$c_{t+1} = r^f \cdot (c_t - \mathbf{1}^\mathsf{T} a_t - \kappa(a_t)) \ .$$

The return rates evolve with the state of the market, as we describe below. The goal of the investor is to maximize a function of the total terminal wealth; it is necessary to not only maximize the expected wealth but also account for the associated risk.

The returns for risky assets evolve according to an exogenous stochastic process that is independent of the investor's position. In particular, the market state is a one-dimensional continuous real variable $z_t$ which partially predicts the returns. The market state and the returns evolve linearly with a jointly normally distributed noise as:

$$\begin{pmatrix} \ln r_{t+1} \\ z_{t+1} \end{pmatrix} = \begin{pmatrix} a_r + b_r z_t \\ a_z + b_z z_t \end{pmatrix} + \begin{pmatrix} e_{t+1} \\ v_{t+1} \end{pmatrix} , \qquad (5)$$

where the stochastic variables $(e_{t+1}, v_{t+1})$ are distributed according to a multivariate normal with zero mean and variance $\Sigma$ for all times. The values $a_r, b_r, a_z, b_z, \sigma$ were estimated from NYSE data for 1927-1996 [11]. Note that $r_{t+1}$ depends only on $z_t$ and is independent of $r_t$. We describe the values of these variables in more detail in Appendix A.

The one-dimensional market state variable and the market rates transitions are discretized [11, 3]. The market state variable is discretized uniformly to 19 points. The distributions for market rates are then estimated for the discrete grid using Gaussian quadrature methods with 3 points per each dimension (27 total) to precisely match the mean and the variance [10].

The dynamic portfolio optimization problem can then be naturally modeled as a hybrid linearly controlled problem. The state space $\mathcal{S} = (\mathcal{D}, \mathcal{X})$ is factored into discrete states $\mathcal{D} = \{1 \ldots 19\}$ that represent the market state $z$ and continuous states $\mathcal{X} \subset \mathbb{R}^4$ that represent the asset investment and the monetary position. The first three elements of $x \in \mathcal{X}$ represent the asset positions and the last element represents the cash.

The set of actions $\mathcal{A}$ represents the feasible trades $a_t \in \mathbb{R}^3$ for each of the three assets. The cash position is adjusted accordingly, depending on the balance of the trades and transaction costs. The actions are constrained by $W(s)$ in order for the asset and cash positions to remain non-negative as follows:

$$W(d, x) = \left\{ a \ : \ \begin{array}{c} x(4) - \mathbf{1}^\mathsf{T} a - \kappa(a) \geq 0 \\ x(i) + a(i) \geq 0 \quad i = 1 \ldots 3 \end{array} \right\} \ .$$

The costs represent the total change in wealth, which is a function of the capital gains and the transaction costs. Formally, for any $(d, x) \in \mathcal{S}$, $a \in \mathcal{A}$, $(d', x') \in \mathcal{S}$ the cost is the reduction in total wealth:

$$c((d, x), a, (d', x')) = \mathbf{1}^\mathsf{T} (x - x') \ .$$

The evolution of the dynamic portfolio optimization is depicted in Fig. 2. The round nodes indicate states, while the square nodes indicate intermediate state-action positions. The solid arrows represent the optimization decisions to rebalance the portfolio and the dashed arrows represent stochastic transitions that determine the next market state and the returns. The figure also indicates at which point the risk mappings are applied.

## 4 Risk Averse Dual Dynamic Programming

In this section, we describe a new approximate algorithm to solve hybrid linearly controlled problems. This algorithm is loosely based on value iteration but uses the polyhedral structure of the MDP to simultaneously update value functions over a range of states. The method is approximate because it uses simulation to identify the relevant states.

The Bellman optimality equations, as described in Theorem 1, can be easily adapted for the hybrid linearly controlled problem as follows.

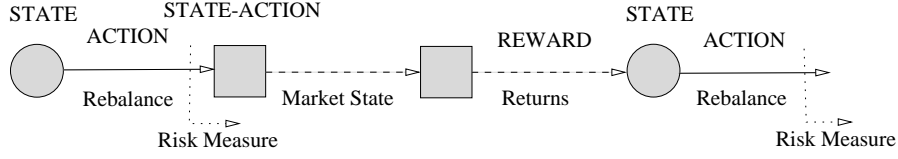

Figure 2: Transitions in the portfolio optimization problem.

**Corollary 1** (Bellman optimality condition). *A finite-horizon value function $v^\star(t)$ is optimal if and only if it satisfies for all $s \in \mathcal{S}$ and $t \in \mathcal{T}$ the following equality:*

$$v_t^\star(s_t) = \min_{a \in W(s_t)} q_t^\star(s_t, a) \qquad (6)$$

$$q_t^\star(s_t, a) = \rho_t\left(c(s_t, a, S_{t+1}) + v_{t+1}^\star(S_{t+1})\right) \qquad (7)$$

*Here, $S_{t+1}$ is a random variable representing the state at time $t+1$ distributed according to the transition probability $P(s_t, a)$ and $v_{T+1}(s_{T+1}) = 0$.*

Assume a state $s = (d, x)$ and consider $D$ and $X$ to be random variables representing the next state as in Definition 1. Then Eq. (7), which is in particular a composition of a risk measure with an affine map, can be written as:

$$\begin{aligned}
q_t^\star(s, a) &= q_t^\star((d, x), a) \\
&= \rho_t\left(c_a^\mathsf{T} a + c_x^\mathsf{T} x + c_n^\mathsf{T} X + v_{t+1}(D, X)\right) \\
&= \rho_t\big(c_a^\mathsf{T} a + c_x^\mathsf{T} x + c_n^\mathsf{T} T_x x + c_n^\mathsf{T} T_a a + c_n^\mathsf{T} U + \\
&\quad + v_{t+1}(D, T_x x + T_a a + U)\big) \\
&= \rho_t\left(Z_{t+1}(d, x, a)\right),
\end{aligned}$$

for the finite vector-valued (random) variable $Z_{t+1}(d, x, a) = c_a^\mathsf{T} a + c_x^\mathsf{T} x + c_n^\mathsf{T} T_x x + c_n^\mathsf{T} T_a a + c_n^\mathsf{T} U + v_{t+1}(D, T_x x + T_a a + U)$, which has dimension $|\Omega_d|$.

To derive RDDP, we now show that the Bellman optimality conditions for linearly controlled hybrid problems can be formulated as a linear program. The optimization variables in the problem correspond to actions and the risk measure. The constraints are a function of the state. Using Eq. (4), it can be readily shown that Eq. (7) a linear program. This representation can then be coupled with the optimization over the action $a$ in Eq. (6) to get the following linear program representation of the Bellman optimality conditions:

$$v_t^\star(d, x) =$$

$$\left[\begin{array}{ll}
\min\limits_{a,\mu,\xi} & (1-\lambda)p_d^\mathsf{T} Z_{t+1}(d, x, a) + \lambda\left(\mu + \frac{p_d^\mathsf{T} \xi}{\alpha}\right) \\
\text{s.t.} & A_d a = b_d - B_d x \\
& l_d \leq a \leq u_d \\
& \xi + \mathbf{1}\mu \geq Z_{t+1}(d, x, a) \\
& \xi \geq \mathbf{0}
\end{array}\right], \qquad (8)$$

where $p_d$ is the measure $P_D$ for $d \in \mathcal{D}$. The notation above assumes that random variables are represented as vectors as the following equality illustrates:

$$p_d^\mathsf{T} Z_{t+1}(d, x, a) = \sum_{\omega \in \Omega_d} p_d(\omega) Z_{t+1}(d, x, a, \omega),$$

where: $\Omega_d$ is the *finite* sample space and $Z_{t+1}(d, x, a, \omega) = c_a^\mathsf{T} a + c_x^\mathsf{T} x + c_n^\mathsf{T} T_x(\omega) x + c_n^\mathsf{T} T_a(\omega) a + c_n^\mathsf{T} U(\omega) + v_{t+1}^\star(D(\omega), T_x(\omega) x + T_a(\omega) a + U(\omega))$.

Now, using Eq. (8), it is easy to show the following proposition that describes the structure of value functions.

**Proposition 1.** *Assuming that $W(s)$ is non-empty for all $s \in \mathcal{S}$, the optimal value function is piecewise affine convex function of $x$ for each $d \in \mathcal{D}$:*

$$v_t^\star(d, x) = \max_{i \in \mathcal{I}_t(d)} \left(q_{x,i}^\mathsf{T} x + q_{c,i}\right), \qquad (9)$$

*where $\mathcal{I}_t(d)$ is a finite non-empty set. In addition, the state-action value function $q_t^\star(s, a)$ is piecewise affine convex in $a$ for any $s \in \mathcal{S}$.*

*Proof.* We prove the proposition by a backward induction on $t$. The proposition holds trivially for $v_{T+1}$ (since $v_{T+1} = 0$). Consider $t = T$. The linear program in Eq. (8) may be simplified as,

$$v_T^\star(d, x) =$$

$$\left[\begin{array}{ll}
\min\limits_{a,\mu,\xi,z} & \lambda \cdot \mu + \sum\limits_{\omega \in \Omega_d} p_d(\omega)\Big((1-\lambda) \cdot z(\omega) + \\
& \quad + \frac{\lambda}{\alpha} \cdot \xi(\omega)\Big) \\
\text{s.t.} & \delta|_1: \quad A_d a = b_d - B_d x \\
& \delta|_2: \quad z(\omega) - (c_a^\mathsf{T} + c_n^\mathsf{T} T_a(\omega)) a = \\
& \qquad = (c_x^\mathsf{T} + c_n^\mathsf{T} T_x(\omega)) x + c_n^\mathsf{T} U(\omega) \\
& \delta|_3: \quad l_d \leq a, \quad \delta|_4: \quad a \leq u_d \\
& \xi(\omega) + \mu - z(\omega) \geq 0, \quad \xi \geq \mathbf{0}.
\end{array}\right] \qquad (10)$$

Due to the assumption of a non-empty and bounded $W(s)$ for all $s \in \mathcal{S}$, the set of extreme points corresponding to the dual polytope of the above linear program is a non-empty, finite set that is independent of $x$, and depends only on $d$. Let $\mathcal{I}_t(d)$ denote this finite set. By duality, we have

$$\begin{aligned}
v_T^\star(d, x) = \max_{i \in \mathcal{I}_t(d)} &\Big(\delta_i|_1^\mathsf{T}(b_d - B_d x) \\
&+ \sum_{\omega \in \Omega_d} \delta_i|_2(\omega)\big((c_x^\mathsf{T} + c_n^\mathsf{T} T_x(\omega)) x + c_n^\mathsf{T} U(\omega)\big) \\
&+ \delta_i|_3^\mathsf{T} l_d + \delta_i|_4^\mathsf{T} u_d\Big)
\end{aligned}$$

The above expression is indeed a piecewise affine convex function of $x$, where we may readily identify $q_{x,i}$ and $q_{c,i}$ by algebraically grouping terms that are linear in $x$ and the additive constant, for each $i$. Thus the proposition is true for $t = T$. Assuming that it is true for any $t + 1 \leq T$, we can show next that it is true for $t$ as well.

Now, for any $t < T$, the mathematical optimization in Eq. (8) with the representation of $v_{t+1}^\star$ as in Eq. (9) is the following linear program.

$$v_t^\star(d, x) =
\begin{bmatrix}
\min_{\substack{a, \mu, \\ \xi, z, y}} \quad \lambda \cdot \mu + \sum_{\omega \in \Omega_d} p_d(\omega)\Big((1 - \lambda) \cdot z(\omega) + \\
\qquad\qquad + \frac{\lambda}{\alpha} \cdot \xi(\omega)\Big) \\
\text{s.t.} \quad \delta|_1 : \ A_d a = b_d - B_d x \\
\quad \delta|_2 : \ z(\omega) - (c_a^\mathsf{T} + c_n^\mathsf{T} T_a(\omega)) a = \\
\qquad\qquad = (c_x^\mathsf{T} + c_n^\mathsf{T} T_x(\omega))x + c_n^\mathsf{T} U(\omega) + y(\omega) \\
\quad \delta|_3 : \ l_d \leq a, \quad \delta|_4 : \ a \leq u_d \\
\quad \delta|_5 : \ y(\omega) - q_{x,j}^\mathsf{T} T_a(\omega) a \geq q_{x,j}^\mathsf{T} \cdot \\
\qquad \cdot (T_x(\omega)x + U(\omega)) + q_{c,j} \\
\qquad\qquad \forall j \in \mathcal{I}_{t+1}(D(\omega)) \\
\quad \xi(\omega) + \mu - z(\omega) \geq 0, \quad \xi \geq \mathbf{0} \ .
\end{bmatrix} \quad (11)$$

The constraint family indexed by $\delta|_5$ is a consequence of the induction hypothesis, where $y(\omega)$ is an auxiliary variable that captures $v_{t+1}^\star(D(\omega), X(\omega))$. A similar argument using duality for Eq. (11) gives:

$$\begin{aligned}
v_t^\star(d, x) = \max_{i \in \mathcal{I}_t(d)} \Big( & \delta_i|_1^\mathsf{T}(b_d - B_d x) + \\
& + \sum_{\omega \in \Omega_d} \delta_i|_2(\omega)\big((c_x^\mathsf{T} + c_n^\mathsf{T} T_x(\omega))x + c_n^\mathsf{T} U(\omega)\big) + \\
& + \delta_i|_3^\mathsf{T} l_d + \delta_i|_4^\mathsf{T} u_d + \\
& + \sum_{\substack{\omega \in \Omega_d, \\ j \in \mathcal{I}_{t+1}(D(\omega))}} \delta_i|_5(\omega, j) \cdot \big(q_{x,j}^\mathsf{T}(T_x(\omega)x + U(\omega))\big) + \\
& + q_{c,j}\Big),
\end{aligned} \quad (12)$$

thereby giving a piecewise affine convex function of $x$ for $v_t^\star(d, x)$. $\square$

Note that the convex representation in Eq. (9) has an exponentially large number of pieces in general, and is not usable in its exact form. The algorithm we propose, RDDP, takes advantage of the piecewise linear representation of the optimal value function to approximate it efficiently. This method is related to value iteration; it iteratively grows an approximation of the large set $\mathcal{I}_t(d)$ by adding one element in each step. Fortunately, it is not necessary to compute the optimal value function to get a good policy—a good value function will suffice. We propose an algorithm that approximates the value function from below by using only a subset of $\mathcal{I}_t(d)$. The algorithm is summarized in Algorithm 1.

In particular, we identify the relevant linear pieces of each value function by successively refining *lower bounds* on the value function. Because the representation as shown in Eq. (9) is convex, a subgradient inequality of the function may be readily derived at any $(d, x)$ which will serve as a lower bound. The following proposition summarizes this property.

**Proposition 2.** *Consider $v_t^\star(d, x)$ described as in Proposition 1, and let $\hat{x} \in \mathcal{X}$ be any fixed continuous component of the hybrid state. For any chosen value $d$, let $\delta_{\hat{i}}$ be the corresponding optimal dual solution of Eq. (11), where $\hat{i} \in \mathcal{I}_t(d)$, and $f^\star(\hat{x})$ be the corresponding optimal objective value. Then: $e_x^\mathsf{T} x + e_c \leq v_t^\star(d, x)$ where:*

$$\begin{aligned}
e_x^\mathsf{T} &= -\delta_{\hat{i}}^\star|_1^\mathsf{T} X_d + \sum_{\omega \in \Omega_d} \big(\delta_{\hat{i}}^\star|_2(\omega) c_n^\mathsf{T} T_x(\omega)\big) \\
&\quad + \sum_{\substack{\omega \in \Omega_d, \\ j \in \mathcal{I}_{t+1}(D(\omega))}} \big(\delta_{\hat{i}}^\star|_5(\omega, j) q_{x,j}^\mathsf{T} T_x(\omega)\big) \\
e_c &= f^\star(\hat{x}) \ .
\end{aligned}$$

*In addition, with such lower bounds added in Algorithm 1, we are guaranteed to converge to the optimal value function.*

*Proof.* The lower bounding inequality for $v_t^\star(d, x)$ follows directly from the subgradient of the convex value function as established in Proposition 1. It is a direct consequence of Eq. (12), since $\hat{i} \in \mathcal{I}_t(d)$. In particular, starting with an empty set $\mathcal{J}_t(d)$ as our approximation for $\mathcal{I}_t(d)$, at each time step we iteratively improve $\mathcal{J}_t(d)$ by adding an additional element $\hat{i}$. The convergence to the optimal solution follows because the scenario tree that represents all $T$-step realizations of the uncertainty is finite. Then, using backward induction, finite termination of the algorithm can be readily shown. $\square$

## 5 Empirical Results

In this section, we present numerical results of RDDP on the portfolio optimization domain described in Section 3. The results not only demonstrate the utility of RDDP in modeling and solving risk-averse MDPs but also illustrate the properties of risk-averse behavior in portfolio optimization. Please note that it is impractical to solve the portfolio optimization problem using discretization since the state-actions space is 8-dimensional. Even with a moderate discretization with 8 points per each dimension, computing the value of the risk measure Eq. (2) for all states and actions

**Algorithm 1:** Risk-sensitive Dual Dynamic Programming (RDDP)

**Data**: MDP Model
**Result**: Lower bound on value function: $\underline{v}_t^\star$
$\mathcal{J}_t(d) \leftarrow \{\} \quad \forall d \in \mathcal{D} \quad \forall t$ ;
**while** *iterations remain* **do**
    // Forward pass: sample states
    $s_0 = (d_0, x_0) \leftarrow$ initial state ;
    **for** $t \in 0 \ldots T$ **do**
        $a_t \leftarrow$ Solve LP Eq. (11) with $\mathcal{J}_{t+1}(d)$ ;
        $\omega_{t+1} \leftarrow$ sample from $\Omega_{d_t}$.
        $x_{t+1} \leftarrow T_x(\omega_{t+1})x_t + T_a(\omega_{t+1})a_t + U(\omega_{t+1})$;
        $d_{t+1} \leftarrow D(\omega_{t+1})$;
    // Backward pass: compute lower bounds
    **for** $t \in T \ldots 1$ **do**
        **for** $d' \in \mathcal{D}$ **do**
            Solve LP Eq. (11) for $(d', x_t)$ ;
            Compute $\hat{i}$, $q_{x,\hat{i}} = e_x$, $q_{c,\hat{i}} = e_c$ from Proposition 2 ;
            $\mathcal{J}_t(d') \leftarrow \mathcal{J}_t(d') \cup \{\hat{i}\}$ ;
    Update lower bound for initial state. Solve LP Eq. (11) for $(d_0, x_0)$ ;

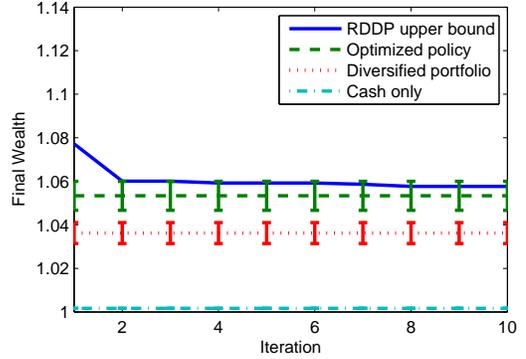

Figure 3: Simulated returns and an upper bound computed by RDDP. The confidence bounds correspond to 2 standard deviations.

involves a linear program with $8^8 * 19 * 27 \approx 9 \cdot 10^9$ variables; here 19 and 27 represent the number of transitions.

We start by evaluating RDDP convergence in a risk-neutral setting; it is easy to evaluate the quality of such a policy by simulation. The policy computed by RDDP for this objective is simple: it always invests the full wealth in the small-cap stock which has the highest expected value. As Proposition 2 shows, RDDP updates a guaranteed lower bound on the optimal value function in each iteration. Because this is a minimization problem, a lower bound on the value function corresponds to an upper bound on the total wealth. Fig. 3 shows the convergence of the lower bound for the consecutive iterations of RDDP. The figure compares the upper bound with simulated returns for 5 time steps and 3000 runs for each policy. The results show that the bound converges close to the optimal solution in 10 iterations. Each iteration involves solving 100 linear programs in Eq. (11).

The opposite of the risk-neutral formulation is the worst-case—or robust—formulation with $\lambda = 1$ and $\alpha = 0$. The robust formulation essentially corresponds to replacing the expectation over the outcomes by the worst case value. The optimal solution in this setting is to invest the entire wealth in cash, which has a constant return and RDDP converges to it in a single iteration. While the worst-case solution is easy to compute, it is hard to evaluate its value by simulation. This is because one needs to sample all realizations to find the worst one.

Finally, we evaluate a mildly risk-averse solution with $\lambda = 0.2$ and $\alpha = 0.7$. Given this objective, the decision maker optimizes the expected return with a weight of 80% and the 70% tail expectation with a weight of 20%. The solution computed for this risk-averse objective invests in a mix of cash and the small-cap stock that depends on the market state. There is no investment in large or medium-cap stock.

Fig. 4 compares the investment value for risk neutral and averse solutions. The investment for the risk-averse solution is 90% in cash and 10% in small-cap stop; the fractions are reversed for the risk-neutral solution. The value of the risk-neutral solution increases with market state because the expected return of stocks increases with higher volatility. The value of the risk averse solution increases for calm market states (1-5) but decreases with higher market volatility as the portfolio mix becomes more biased towards cash. The value function the decreases sharply when the volatility is high because of trading fees charged when rebalancing towards cash.

The risk-averse solution achieves an average return of about 0.6%, which is much lower than 5.4% for the risk neutral solution. However, the variance of the risk-averse solution is close to 0 and the return is about 5 times greater than the return of pure cash investment 0.1%. RDDP also efficiently optimizes the dynamically consistent risk measure reducing the lower bound by a factor of 26.9 in the first three iterations.

It is interesting to compare the RDDP solution to existing methods for solving portfolio optimization problems. The risk aversion in previous work was modeled by a concave utility function and the solutions were heuristics based on frictionless models [11, 3].

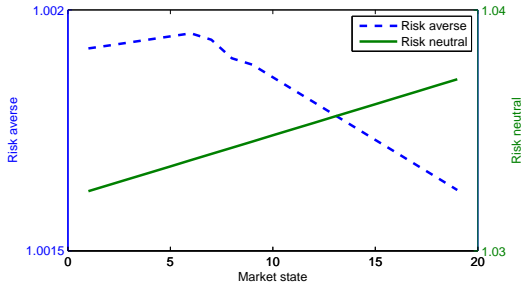

Figure 4: Value of investment as a function of the market state (negated value function). Market volatility increases with increasing state.

| | Large-cap | Mid-cap | Small-cap | Cash |
|---|---|---|---|---|
| Mean ($a_r$, $a_z$) | 0.0053 | 0.0067 | 0.0072 | 0.0000 |
| Coefficient ($b_r$, $b_z$) | 0.0028 | 0.0049 | 0.0062 | 0.9700 |

Table 1: Regression coefficients of rate means

| | Large-cap | Mid-cap | Small-cap | Cash |
|---|---|---|---|---|
| Large-cap | 0.002894 | 0.003532 | 0.003910 | -0.000115 |
| Mid-cap | | 0.004886 | 0.005712 | -0.000144 |
| Small-cap | | | 0.007259 | -0.000163 |
| Dividend Yield | | | | 0.052900 |

Table 2: Noise covariance $\Sigma_{ev}$

The utility-based risk aversion, coupled with the assumption of no transaction fees, led to diversification of the portfolio between middle and large capitalization stocks. On the other hand, the RDDP solution with the AV@R risk measure prefers diversification into cash to other stocks. Future work will need to address whether this difference is due to different risk objectives or different approximation techniques used to compute the solutions.

## 6 Related Work and Discussion

In this section, we discuss connections between RDDP and related work in both stochastic programming and artificial intelligence communities. Large or continuous risk-neutral MDPs, such as the hybrid linearly controlled problems, are often solved by approximate dynamic programming or reinforcement learning. Approximate dynamic programming overcomes the large size of these problems by restricting value functions to a small linear space [18]. We are not aware, however, of any approximate dynamic programming methods that optimize risk-averse objectives. RDDP is similar to approximate dynamic programming, except the features are constructed automatically using the special structure of the problem. As a result, RDDP is easier to apply to compatible problems, but is less general because it cannot be applied to non-linear problems.

RDDP is also related to stochastic dual dynamic programming (SDDP). SDDP is an approximate method for solving large multistage linear programs with certain independence assumptions [14, 15, 22]. Some of these methods have been extended to risk averse objectives [9, 6, 17, 7]; although they typically assume a different independence structure from the proposed linearly controlled problems.

In a parallel stream of work on risk-averse optimization, models with polyhedral risk measures have been studied [4, 8]. Unlike the Markov risk measures, polyhedral risk measures are not dynamically consistent, but must satisfy other properties that make the resulting problems easy to solve.

## 7 Conclusion

The paper describes RDDP, a new algorithm for solving Markov decision processes with risk averse objectives and continuous states and actions. We model the risk-aversion using dynamically consistent convex risk measures, which is an established approach in stochastic finance. Our experimental results on a portfolio optimization problem indicate that RDDP can be converge quickly to a close-to-optimal solution for both risk-averse and risk-neutral objectives. Our risk averse solutions also provide new insights into the properties of risk-aversion in the portfolio optimization setting.

While the focus of this paper is on a financial application, there are numerous other domains that exhibit similar linear properties, such as stochastic inventory management, or hydrothermal energy management. The portfolio optimization problem represents a good benchmark, because it is relatively simple to describe and fit to real data, and a similar structure can be found in many domains that involve optimal utilization of finite resources.

One significant weakness of RDDP is that the convergence criterion is not well defined and risk-averse objectives may be hard to evaluate by simulation. It is not hard, however, to remedy both these issues. Using the convexity of the optimal value function and the lower bound computed by RDDP, it is also possible to compute an upper bound on the value function based on Jensen's inequality. The difference between the upper and lower bounds then provides a reliable stopping criterion and an empirical value of a policy.

## A Numerical Values

The values $a_r, b_r, a_z, b_z, \sigma$ were estimated from NYSE data for 1927-1996 [11]. Table 1 summarizes the regression coefficients $a_r, b_r, a_z, b_z$ and Table 2 summarizes the covariance of the noise $\sigma$. The risk-free return on cash is $r_f = 1.00042$. The initial market state is $z_0 = 0$.